%% file: main.tex
\documentclass[runningheads]{llncs}
\input{config}
\input{mymacros}

\usepackage[utf8]{inputenc}
\usepackage[T1]{fontenc}
\usepackage{xspace}
\usepackage{stmaryrd}
\usepackage{./why3lang}
\usepackage{./gospel}
\usepackage{./lstcoq}
\usepackage{color, colortbl}
\usepackage{graphicx}
\usepackage{array}
\usepackage{cellspace}
\usepackage{cite}
\usepackage{latexsym}
\usepackage{subcaption}
\usepackage{hyperref}
\usepackage{cleveref}

\def\_{\kern.08em\vbox{\hrule width.35em height.6pt}\kern.08em}

\newcommand{\nbnote}[3]{
  \fcolorbox{gray}{yellow}{\bfseries\sffamily\scriptsize#1}
  {\color{#2} \sffamily\small$\blacktriangleright$\textit{#3}$\blacktriangleleft$}
}

\newcommand{\mario}[1]{\nbnote{Mario}{red}{#1}}
\renewcommand{\mario}[1]{}
\newcommand{\ion}[1]{\nbnote{Ion}{blue}{#1}} %
\renewcommand{\ion}[1]{}%
\newcommand{\tiago}[1]{\nbnote{Tiago}{green}{#1}} %
\renewcommand{\tiago}[1]{}%

\begin{document}
\title{Static and Dynamic Verification of \ocaml Programs: The
  \gospellang Ecosystem\\(Extended Version)}
\titlerunning{Static and Dynamic Verification of \ocaml Programs: The
  \gospellang Ecosystem}
\author{Tiago Lopes Soares \and Ion Chirica
  \and Mário Pereira}
\institute{NOVA LINCS, Nova School of Science and Technology, Portugal}

\maketitle
\begin{abstract}
  We present our work on the collaborative use of dynamic and static analysis
  tools for the verification of software written in the \ocaml language. We
  build upon \gospellang, a specification language for \ocaml that can be used
  both in dynamic and static analyses.
  We employ \ortac, for runtime assertion checking, and \cameleer and \cfmllang
  for the deductive verification of \ocaml code. We report on the use of such
  tools to build a case study of collaborative analysis of a non-trivial \ocaml
  program. This shows how these tools nicely complement each others, while at
  the same highlights the differences when writing specification targeting
  dynamic or static analysis methods.

\end{abstract}


\section{Introduction}
\label{sec:intro}

In the past decades, we have witnessed remarkable developments in the field of
formal methods. In particular, Deductive Software
Verification~\cite{filliatre11sttt} and Runtime Assertion Checking
(RAC)~\cite{10.1145/1127878.1127900} tools have evolved into practical,
scalable, and trustworthy systems that can be used to verify complex pieces of
software. However, with some notable exceptions,
% análise dinâmica VS análise estática
dynamic and static analysis are very rarely used collaboratively to analyze
(parts of) a program.
% normalmente utiliza-se um ou outro
% Formal methods practitioners tend to adopt either or the other: dynamically
% check a system, at the expense of less expressiveness and not being complete;
% or statically analyze a program, at the cost of more human interaction hence less
% automation.
Formal methods practitioners tend to choose either dynamic or static
analysis, each with its own limitations: dynamic checks sacrifice expressiveness
and completeness, while static analysis requires more human interaction, hence
it is less automated.  \mario{Não estou muito satisfeito com esta frase, se
  calhar tiramos da versão final do abstract/introdução.}
% nosso argumento: a correcta utilização seria uma utilização colaborativa
% deste tipo de ferramentas e abordagens
We strongly believe this separation is inherently restrictive and advocate for
the collaborative use of dynamic and static tools to analyze the same
software system.

% a nossa proposta neste artigo
In this paper, we present our work on the joint use of dynamic analysis and
static verification tools. Specifically, we focus on programs written in the
\ocaml language.  \tiago{em relação ao texto comentado, estas abordagens são
  cooperativas? Analise dinamica facilita a estatica, mas acho que não acontece
  ao contrario (pelo menos, não falamos disso no artigo :|).}
% utilização conjunta e colaborativa
% Our vision is that dynamic and static analysis should co-exist and cooperate
% naturally to certify a piece of software. In particular, the \emph{formal
% specification} of a system should be (as much as possible) interchangeable
% and reusable between the two approaches.
In our approach, static and dynamic analysis are treated not as mutually
exclusive but rather as two sides of the same coin. Our vision is that these two
techniques can be combined by operating over (roughly) the same
specification. \tiago{Nova versão deste paragrafo}%
\mario{Retirei a parte do ``novel'', pois já temos (pelo menos) os exemplos de
  SPARK (linguagem ADA) e Frama-C (linguagem C) em que existe colaboração entre
  dynamic e static.}
% uma linguagem de spec capaz de expressar ambas as técnicas
A core ingredient in our approach is a specification language usable by both
analysis approaches.
% \gospel
We
% make use of %leverage on
leverage \gospellang~\cite{ChargueraudFLP19}, a tool-agnostic \ocaml
specification language that serves as a common ground for the communication
between dynamic and static methods. \gospellang is strongly inspired by other
behavioral specification languages~\cite{DBLP:journals/csur/HatcliffLLMP12},
namely \textsf{SPARK}~\cite{DBLP:conf/sigada/CarreG90},
\textsf{JML}~\cite{DBLP:journals/sigsoft/LeavensBR06}, and
\textsf{ACSL}~\cite{acsl_reference} which can also be used both for dynamic and
static analysis of code.

% our pipeline
\mario{Esta ideia do pipeline poderia vir só depois da indicação da utilização
  do \gospellang}%
We propose the following, rather natural, certification workflow: first, use RAC
tools to increase the confidence on the devised specification for parts of a
program; second, apply deductive verification on that very same piece of code to
achieve even higher correctness guarantees.
% ferramentas
Our analysis pipeline is materialized, on the one hand, through the
use of \ortac~\cite{DBLP:conf/rv/FilliatreP21}, a RAC tool for \ocaml,
and on the other hand through the deductive verification of \ocaml
code using \cameleer ~\cite{pereiracav21} and
\cfmllang~\cite{cfml}. All three are able to interpret and process
\gospellang specifications.

% diferenças, pros e cons
We apply our certification pipeline to a piece of \ocaml code, issued
from the widely used \textsf{OCamlGraph} library, whose verification
is non-trivial. We apply RAC to parts of the program, namely auxiliary
data structures, and then incrementally use deductive verification to
achieve higher correctness guarantees. This shows how the different
tools of the \gospellang ecosystem complement each others, but also
highlights the differences of writing \gospellang specifications
targeting dynamic or static analysis.

% perspectivas interessantes na aplicação de métodos formais na comunidade
We believe our proposal opens interesting perspectives on the practical adoption
of formal methods among the \ocaml %programmers
community. To the best of our knowledge, this is the first time a collaborative
use of RAC and deductive verification tools is done in the setting of
multi-paradigm language like \ocaml. Finally, we believe our reported experience
makes a contribution towards bridging the gap between different program
specification and analysis paradigms.

This paper is organized as follows. In Section~\ref{sec:gospel-eco},
we survey the \gospellang language and the tools that compose its
ecosystem. In Section~\ref{sec:cert-workfl}, we describe our proposal
certification workflow for the collaborative use of dynamic and static
analysis of \ocaml code. In Section~\ref{sec:path-checking-graph}, we
put our pipeline to work on the case study. Finally, we conclude with
some related work, Section~\ref{sec:related-work}, and closing remarks
and future perspectives in Section~\ref{sec:concl-future-work}. All
the software and examples used in this paper are publicly available in
a companion artifact~\cite{soares-isola2024}.

% present an elaborated example of our
% pipeline in practice. We test some parts using \ortac ... we then rely on the
% tested specification to conduct the proof, in \cameleer, of the main algorithm
% ... \cfmllang is used to close up the proof ...

\mario{``bridge across different specification and verification techniques,
  including dynamic and static analysis,'', cf
  \url{https://2024-isola.isola-conference.org/isola-tracks/}}

\mario{É nesta frase da apresentação da track que me estou a basear. No abstract
  ou na introdução devemos voltar a isto, até mesmo utilizar algumas destas
  expressões.}

\mario{Dizer que o \gospellang é inspirado noutras linguagens, que até já foram
  utilizadas para RAC e Deductive Verification. E.g., JML and ACSL}

\mario{Dizer algures na introdução que queremos que a nossa abordagem seja um
  esforço colaborativo: podemos querer apenas testar algumas coisas; provar
  outras assumindo a spec que testamos.}

\mario{Insistir que as diferentes análises requerem, em alguns casos, diferentes
  formas de especificação. Tentamos colmatar essas diferenças. Devemos voltar a
  este ponto na conclusão.}

% ~\Cref{sec:path-checking-graph} presents a more elaborated example of our
% pipeline ... We test some parts using \ortac ... we then rely on the tested
% specification to conduct the proof, in \cameleer, of the main algorithm
% ... \cfmllang is used to close up the proof ...

\paragraph{Labels on code listings.} Throughout this paper, we present
many different code listings written in the three languages our tools
manipulate, \emph{i.e.}, \gospellang, \ocaml, and \cfmllang (which is
basically \coq syntax augmented with Separation Logic operators). To
ease the reading process, we attach a label to each listing, on the
top-right corner, that indicates the input language(s).

\section{The \gospellang Ecosystem, in a Nutshell}
\label{sec:gospel-eco}

\subsection{Gospel -- A Specification Language for \ocaml}
\label{sec:gospel}

\mario{Ainda não li esta secção, mas precisamos de introduzir além dos
  \of{requires}, \of{ensures} e \of{modifies} a cláusula \of{raises}. Esta vai
  ser usada na especificação \ortac da \texttt{Queue}}

\gospellang~(Generic OCaml SPEcification Language) is a behavioral specification
language for \ocaml code. It is a contract-based, strongly typed language, whose
semantics is defined in terms of a translation into Separation
Logic~\cite{sep-log}. \gospellang terms are written in a subset of \ocaml,
augmented with quantifiers.
Although \gospellang specifications are semantically equivalent to
Separation Logic, these are much more lightweight and concise to read
and write. For instance, \gospellang assumes by default that function
parameters, as well as their return values, are separated in memory
and that the caller has full ownership. This eliminates the need for
explicitly stating properties that would seem natural to an everyday
programmer, but nevertheless would be necessary in other platforms
built on Separation Logic, such as VeriFast~\cite{verifast} or
Viper~\cite{viper}.

\begin{figure}[t]
  \centering
  \begin{gospelsmall}
  type 'a t
  (*@ mutable model elems : 'a list *)

  val create : unit -> 'a t
  (*@ q = create ()
        ensures q.elems = [] *)

  val push : 'a t -> 'a -> unit
  (*@ push q x
        modifies q
        ensures q.elems = old q.elems @ [x] *)

  exception Empty

  val pop : 'a t -> 'a
  (*@ x = pop q
        modifies q
        raises Empty -> old q.elems = [] = q.elems
        ensures x :: q.elems = old q.elems *)

  val is_empty : 'a t -> bool
  (*@ b = is_empty q
        ensures b = q.elems = [] *)

  val transfer : 'a t -> 'a t -> unit
  (*@ transfer q1 q2
        modifies q1, q2
        ensures q1.elems = [] && q2.elems = old (q2.elems @ q1.elems) *)
    \end{gospelsmall}
  \caption{\gospellang Queue Specification.}
  \label{fig:queue}
\end{figure}

In \Cref{fig:queue}, we present an \ocaml interface, adapted from the
original \gospellang paper~\cite{ChargueraudFLP19}, where we specify
the behaviour of a polymorphic FIFO queue. We start by defining the
queue type \gosp{'a t} and giving it a \gosp{model} field. A
\gosp{model} is a field that can only be accessed in specifications
and acts as a logical representation of the type. In this case, the
logical representation of the queue is a \gosp{'a list}. By marking
this field as \gosp{mutable}, we state that the contents of the queue
maybe mutated in-place.

The \gosp{create} function returns an empty queue, the \gosp{push} and
\gosp{pop} functions modify the model of the queue by inserting at its
tail and removing from its head, respectively. Additionally,
\gosp{pop} also throws an exception if the queue is empty, not
changing the contents of the model. The \gosp{is_empty} function
decides whether the model of the queue is the empty list. Finally,
\gosp{transfer} empties \gosp{q1} and concatenates all its elements to
the model of \gosp{q2}. Its specification also implicitly states that
both queues are separate in memory. To showcase the benefits of this
design choice, we present an equivalent Separation Logic specification
where we assume the existence of a representation predicate $Q$ that
relates a memory location with its logical representation (in this
case, a list) as well as claiming ownership of the data structure:
\begin{align*}
  \forall &q_1\ q_2\ L_1\ L_2,\\ &\{Q(q_1, L_1)\Qstar Q(q_2, L_2)\}
  (\texttt{transfer}\ q_1\
  q_2) \{Q(q_1, nil) \Qstar Q(q_2, L_2 \texttt{++} L_1)\}
\end{align*}
By writing our specifications in \gospellang, we avoid having to
explicitly write conditions relating to ownership and separation and
can focus on describing the observable behaviour of the function. It
is worth noting that \gospellang is not tied to any particular tool or
analysis framework. The remaining of this section presents the tools
one can use to interface with \gospellang, in order to dynamically and
statically verify \ocaml programs.

\subsection{Ortac -- Runtime Assertion Checking of \ocaml Programs}
\label{sec:ortac}

\mario{Dar algures o link para o Github do \ortac?}

\ortac is a RAC tool for \ocaml code. It consumes
\gospellang-annotated \ocaml interface files to generate a wrapper
around a candidate implementation that dynamically checks the supplied
specification. This wrapper uses the implementation as a black-box: it
tests the validity of the precondition when the function is called and
check if the postcondition holds when the function returns.
% Such a wrapper works in a black-box fashion, not modifying the implementation
%  but simply testing the specification at function call and
% return. In particular, it tests the validity of preconditions for the interface
% exposed functions, and also that postconditions hold when a function returns.

The design and use of the \ortac framework are centered around two main
principles: first, it should work fully automatically, only requiring a user to
write the \gospellang specification; second, it is a \emph{modular} and
\emph{extensible} tool, where users can extend the behaviour and capabilities of
\ortac via \emph{plugins}. The former is an important requirement for \ocaml
programmers to smoothly integrate RAC in development pipelines, where \ortac can
be used as an additional guarantee layer on deployed code.
The latter is a crucial feature of the tool, since it provides the
necessary building blocks to accommodate different specification-based
testing techniques around a core \ortac. In its core, \ortac features
a translation mechanism from the \gospellang executable subset into
\ocaml code. Each plugin exploits this translation to implement
different analysis strategies of \gospellang specifications.

Two main plugins have already been developed for \ortac:
(i) an interface to \textsf{Monolith}~\cite{pottier2021strong}, a fuzz testing
  tool for \ocaml libraries;
(ii) the \stm plugin, for model-based testing of \ocaml
  implementations.
In this paper, we frame our use of \ortac to the \stm option. The main purpose
of this plugin is to generate a set of random calls to functions exposed in an
\ocaml interface file and checks whether their operational behaviour matches the
logical behaviour expressed in \gospellang annotations. From a practical point
of view, this plugin wraps \gospellang specification and a candidate
implementation file into instrumented code that interfaces with
\textsf{STM}~\cite{midtgaard2022multicoretests}, a state-machine testing library
for \ocaml code. In turn, \textsf{STM} builds on \textsf{QCheck}~\cite{qcheck},
an \ocaml property-based testing framework inspired by
\textsf{QuickCheck}~\cite{DBLP:conf/icfp/ClaessenH00}.
% The goal is to use
%\ortac to test an \ocaml implementation of queues against such \gospellang
%specification.

% The model of the queue is now
% an \ocaml list instead of a mathematical sequence (the model must be an
% executable \ocaml structure), with the sequence operations being replaced by
% equivalent list operations. For the sake of completeness, we must now test the
% \of{pop} function for the case where the queue can actually be empty, hence this
% function can now raise the \texttt{Empty} exception. Also, one must provide a
% postcondition of the form \of{t.elems = ...} to capture the state of the model
% when the function returns, as well as a separated postcondition to assert how
% the returned value relates to the model. Finally, the \stm plugin is only able
% to analyze functions featuring a single argument of type \of{'a t}, the type of
% queues, hence we remove the \of{transfer} function. The adapted \texttt{Queue}
% specification, used by \ortac, is given in Appendix~\ref{sec:queue-spec-ortac}.

\mario{Argumentar, informalmente, que as specs são equivalentes?}

\tiago{Não argumentei, mas pus uma pequena frase a dizer que a spec em anexo e a
spec da figura são equivalentes}

\mario{Dizer aqui que queremos a maior parte do tempo utilizar especificação
  baseada em funções da própria biblioteca standard \ocaml, para podermos
  executar essa spec.}

\subsection{Cameleer -- Auto-active Verification of \ocaml Programs}
\label{sec:cameleer}

\mario{dizer que o \cameleer é state of the art é talvez too much}

\cameleer is a tool for deductive verification of \ocaml programs,
conceived with proof automation in mind. It takes as input
\gospellang-annotated \ocaml code and translates it into an equivalent
\whyml representation, the programming and specification language of
the \whythree framework~\cite{why}. One of the key strengths of using
\whythree is its ability to interface with several theorem provers,
whether automated or interactive, ultimately providing a more flexible
and ergonomic proof experience. Furthermore, \whythree allows users to
conduct some lightweight interaction~\cite{dailler2018} to aid in the
proof of a failing Verification Condition (VC).

To introduce \cameleer, let us consider the queue data structure and
the implementation of the \gosp{push} operation. A queue is defined
using two lists, \gosp{front} and \gosp{rear}, as follows:
\begin{gospelsmall}
  type 'a t = {
    mutable front : 'a list;  mutable rear : 'a list;
  }
  (*@ mutable model elems : 'a list *)
  (*@ with x invariant (x.front = [] -> x.rear = []) &&
                        x.elems = x.front @ List.rev x.rear *)
\end{gospelsmall}
Elements are pushed into the head of the \gosp{rear} list and popped
from the \gosp{front} list.
This type is also given a type invariant, which states that if
\gosp{front} is empty, so is the \gosp{rear}, and that the model
\gosp{elems} stands for the concatenation of \gosp{front} and the
reverse of \gosp{rear}. If the queue is empty then we assign
\gosp{front} the singleton list~\of{[x]}; otherwise we add it to the
head of \gosp{rear}. Additionally, we update \gosp{elems} by appending
the element at the tail, as follows:
\begin{gospelsmall}
  let push x q =
    if is_empty q then q.front <- [x] else q.rear <- x :: q.rear;
    q.elems <- q.elems @ [x]
  (*@ push x q
      modifies q
      ensures q.elems = old q.elems @ [x] *)
\end{gospelsmall}
The specification of \gosp{push} is given with regards to
\gosp{elems}, the \gosp{model} of the queue. Note how this \gospellang
specification is the same as the one in the interface shown
in~\Cref{fig:queue}.
Assuming the file \of{queue.ml} contains the above implementation,
starting a proof with \cameleer is a simple matter of calling the
following command: \texttt{cameleer queue.ml}.
\cameleer translates the input program into an equivalent \whyml representation,
launching a graphical interface for the \whythree IDE.
%
% The \whythree graphical interface allows us to individually prove VCs
% (Verification Conditions), by decomposing a compound task into smaller, more
% digestable VCs --- this can be done with \texttt{split\_vc} in the dedicated
% command line.  Irrespective of how big the task is, either a compound or an
% irreducible VC, we can dispatch external provers by selecting a VC and
% explicitly choose a specific prover in \menu[,]{Tools, Provers}, or
% collectively call every available prover in \menu[,]{Tools, Strategies}. In
% this case,
%
For the implementation of all queue operations, \whythree generates a
total of 40 VCs, all quickly discharged by the Alt-Ergo~\cite{alt} SMT
solver.

\whythree is a very natural choice to translate \gospellang into,
given that they are, at least on a surface level, very
similar. Nevertheless, its semantics differ in that \whythree does not
employ Separation Logic or any kind of permission based logic. Even
though the \whythree type-and-effects system makes similar assumptions
relative to what is expected in
\gospellang~\cite{filliatre2016pragmatic}, it falls short when it
faces programs with more sophisticated constructs such as recursive
types with mutable fields.

\mario{Verificar onde colocar esta citação:~\cite{dailler2018}}
\tiago{Adicionar um paragrafo com as limitações do \cameleer para
  haver uma transição entre as duas secções.}

\subsection{CFML -- Interactive Verification of \ocaml Programs}
\label{sec:cfml}

In order to fully capture the meaning of \gospellang specifications we
must turn to a more expressive logic. We choose \cfmllang
(Characteristic Formulae for ML), a \coq~\cite{coq} framework created
for the verification of \ocaml code using Separation Logic. \cfmllang
is composed of:
\begin{itemize}
\item A translator mechanism, which takes an \ocaml program
  and generates a \coq translation of such program;
\item A higher-order Separation Logic encoding in \coq, together with
  a comprehensive library of tactics.
\end{itemize}
To verify \gospellang with \cfmllang, we first use the aforementioned
translator to translate the \ocaml implementation into \coq. Next, we
translate the \gospellang specifications into \cfmllang by means of a
prototype
tool\footnote{\url{https://github.com/ocaml-gospel/gospel2cfml}}. Finally,
we write a proof script proving the generated specification adheres to
the \ocaml implementation.

To demonstrate how we deal with \gospellang in \cfmllang, we feed it
the specification in \Cref{fig:queue}. In the case of such an
interface, we first translate the type declaration. Since the type has
a mutable \gosp{model} field it is ephemeral, meaning it is treated as
a \of{loc} value, the \cfmllang type for mutable \ocaml values. To
reason about a queue, we assume a representation predicate that
relates its entry pointer and a logical list of its elements. This is
as follows:
%
% encoded directly into \coq. Therefore, to talk about its
% model we will assume a representation predicate that lifts a
% \gosp{loc} (CFML's type for mutable \ocaml values) into a \gosp{list
%   A}.
%
\begin{cfml}
  Parameter Queue : forall A, list A -> loc -> hprop.
\end{cfml}
Note how the return value of this representation predicate is not
\gosp{Prop}, but rather \gosp{hprop}, the type of heap-parameterized
propositions. Our tool does not generate the body for this predicate,
it must be defined later by the user once an implementation for the
\of{queue} type is provided.

What remains to be translated are the \ocaml function declarations,
together with \gospellang contracts. Let us take \of{push} as an
example. A top-level function is represented in \cfmllang using the
\of{val} type:
\begin{cfml}
  Parameter push : val.
\end{cfml}
The \gospellang specification of the \of{push} function translates as
follows:
\begin{cfml}
  Lemma push_spec :
    forall A (q : loc) (elems : list A) (x : A),
    SPEC (push q x)
      PRE (q ~> Queue elems)
      POSTUNIT (\exists elems', q ~> Queue elems' \* [elems' = elems ++ [x]]).
\end{cfml}
As mentioned in \Cref{sec:gospel}, \gospellang makes several
assumption in regards to ownership of values. In this case, it assumes
that the \gosp{queue} is owned before and after a call to \gosp{push}.
Naturally, these properties must now be made explicit in the \cfmllang
specification.

In the precondition, ownership of \gosp{q} is claimed, and its model
is now accessible. Notation \gosp{q ~> Queue elems} stands for a
well-formed queue with entry pointer~\of{q}, whose model is
\of{elems}.
In the postcondition, we once again claim ownership of \gosp{q} while
also stating that it is now represented by the updated model
\gosp{elems'}, which is introduced using \of{\exists}, the Separation
Logic existential quantifier. The \gospellang postcondition,
\emph{i.e.}, the pushed element is at the tail of the updated list, is
encoded as well. This condition is within a \coqinline{[...]}  block,
meaning it is a pure assertion, \emph{i.e.}, it does not dependent on
heap-allocated elements.

% Next, we translate the \gosp{push} function. Before getting into the
% specifications, we must first assume that there are values that
% represent these \ocaml functions in our \cfmllang file. These values
% will have type \gosp{val}, the type of non-mutable \ocaml values in
% \cfmllang. This value, like our representation predicate, will remain
% undefined and will be later actualized by \cfmllang's translation of
% the implementation.

% With this, we can now translate the \gospellang specification into
% \cfmllang.

The remaining generated \cfmllang specifications are relatively
similar: for each \ocaml top-level function it creates a
\of{Parameter} that represents such a function, claiming ownership of
the mutable parameters via the \of{Queue} representation
predicate. All user-supplied \gospellang specification are translated
into pure blocks.

\section{Certification Workflow}
\label{sec:cert-workfl}

\mario{Será ``workflow'' ou ``pipeline''? Talvez ``workflow'' no título esteja
  bem aqui, mas durante o texto podemos ir trocando entre ``workflow'' e
  ``pipeline''.}

\mario{Ainda não reli esta secção. Mas algo que precisamos de colocar em ênfase
  é que o nosso pipeline não deve obrigar a que \textbf{tudo} seja testado e
  \textbf{tudo} seja verificado. Devemos indicar claramente que é uma escolha do
  utilizador onde quer empregar cada ferramenta.}
\mario{Até porque isso é importante para o nosso exemplo grande, já que testamos
  apenas as estruturas de dados auxiliares num primeiro tempo e confiamos nessa
  spec para fazer a prova da função \gosp{path_check}. Só num segundo tempo é
  que decidimos verificar formalmente as estruturas de dados auxiliares.}

\begin{figure}[t]
  \centering
  \includegraphics[width=0.57\linewidth]{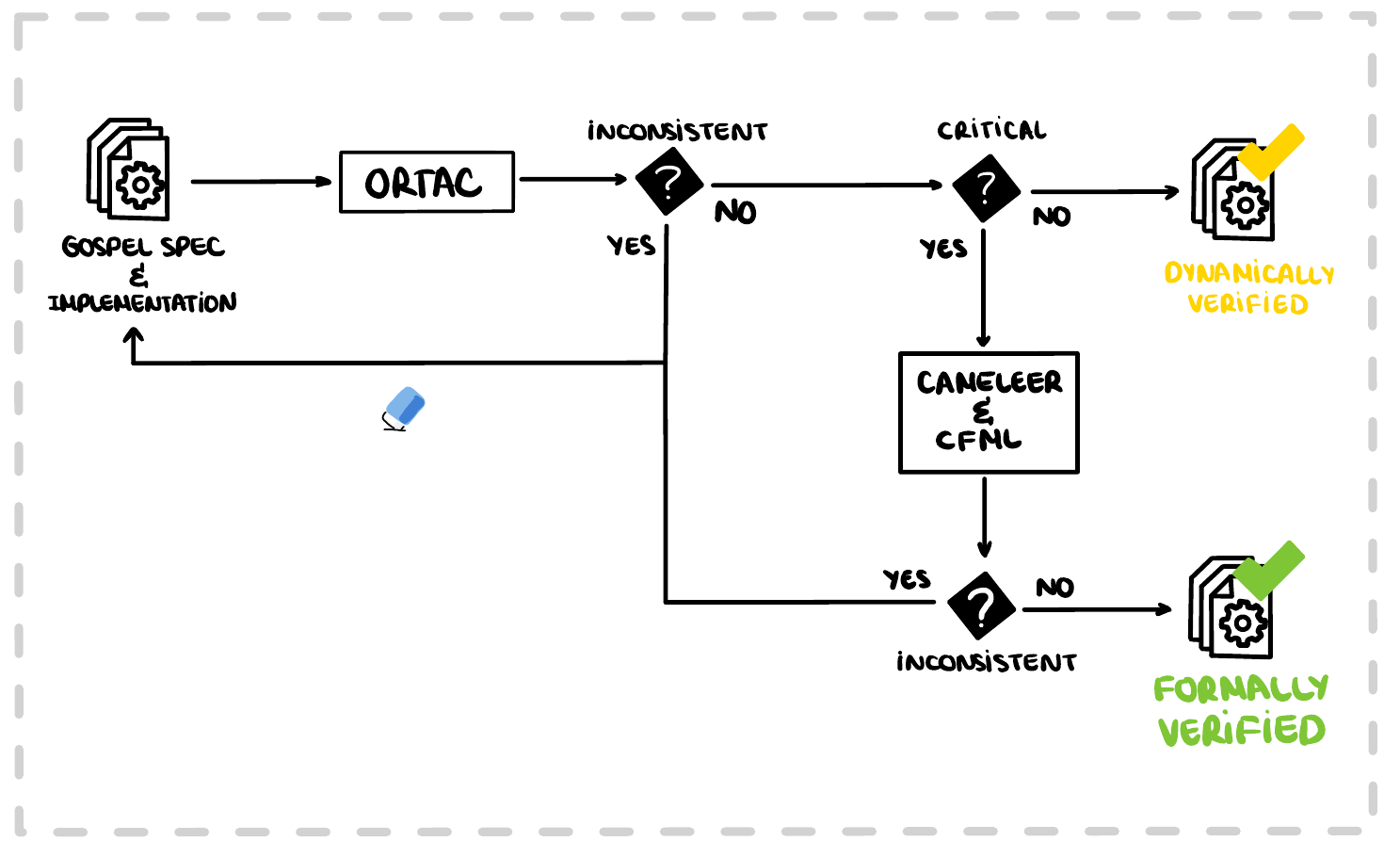}
  \caption{\gospellang certification workflow.}
  \label{fig:pipeline}
\end{figure}

In this section, we present our proposal for a certification workflow
that combines the tools of the \gospellang ecosystem. We show how one
can employ different analysis paradigms to achieve a flexible and
adaptable pipeline. Our proposal is graphically summarized in
\Cref{fig:pipeline}.

% Now that we have gone over each specific link in the \gospellang verification
% toolchain, we will now see how they complement one another to form a coherent
% certification workflow.

The core element in our pipeline are \gospellang-annotated \ocaml
interface files. Such interfaces must capture the behavior of parts of
the whole implementation, meaning the user might only wish to analyze
specific points in their code.
Together with such interface files, one must provide \ocaml
implementation for the specified functions.

As a first analysis step, we propose to use \ortac to dynamically
check the implementation adheres to the supplied specification. This
is a natural entry point in our analysis diagram, since \ortac is the
tool that requires less user interaction, hence easier to include in a
development pipeline.

At this point, one can very well be satisfied with the assurances that
\ortac provides. However, if the programmer requires a greater degree
of confidence in the correctness of their code, they can formally
verify (parts of) their programs using \cameleer and/or \cfmllang,
once \ortac can no longer find mismatches between code and
specification.  It is important to note that the effort to prove the
correctness of a program should only begin once we are almost certain
that the specification holds. However, there are situations where the
proof will fail either due to an incomplete specification or some
problem with the code itself. In both cases, and especially when many
changes are made, it is beneficial to run the updated program through
\ortac, to check if the modifications did not break our program.

Between the two tools, \cameleer is easier to work with since the SMT
solvers do the heavy lifting of discharging the necessary verification
conditions. On the other hand, \cfmllang is the tool of choice when it
comes to reason about more complex \ocaml programs, in particular
those that deal with heap-allocated data structures.

\section{Case Study: Path Checking in a Graph}
\label{sec:path-checking-graph}

\mario{Ainda não reli esta secção. Mas um aspecto importante em que é preciso
  insistir é o facto desta função \gosp{check_path} funcionar independentemente
  da implementação concreta do grafo e dos nós. Por outras palavras, insistir no
  carácter modular desta implementação.}

As our main case study, we chose an algorithm adapted from
\ocamlgraph~\cite{ocamlgraph}, a library featuring a large set of
generic graph data structures and operations. This algorithm checks
whether there is a path in a graph between two given nodes. In this
section we first present the implementation that will be used
throughout, followed by a dynamic analysis on \ortac; afterwards we
present a proof of the algorithm in \cameleer; and finally, a proof in
\cfmllang.

\mario{Roadmap da secção. Dar a entender as diferentes etapas que
  vamos explorar aqui.}

% Though we omitted the optimization of caching paths that are already
% computed, we believe that its specification and subsequent proof are
% not greatly affected by this relaxation. We find it to be an
% interesting candidate for our case study, seeing as it makes use of
% the \texttt{Queue} we specified in \Cref{sec:cfml}.

\subsection{Implementation}
\label{sec:implementation}

The path checking algorithm is implemented as a functor \of{Check},
making it completely modular to the actual implementation of the graph
data structure and operations.
In this case, the functor argument is a module that contains the type
of graphs, \of{gt}, and a function that returns the list of successors
for some vertex. Additionally, graph vertices must respect the
\of{COMPARABLE} module signature which defines a type equipped with
order and equivalence relations and a hash function. For the sake of
completeness, the definition of \of{COMPARABLE} can be found
in~\Cref{sec:vertex}.

% \begin{gospelsmall}
%   module Check
%     (G : sig
%       module V : COMPARABLE
%       type gt
%       (*@ model dom: V.t fset
%           model suc: V.t -> V.t fset
%           with x invariant forall v1, v2.
%              mem v1 x.dom -> mem v2 (x.suc v1) -> mem v2 x.dom *)

%       val [@logic] successors : gt -> V.t -> V.t list
%       (*@ l = successors graph source
%           requires mem source graph.dom
%           ensures forall v'. List.mem v' l <-> mem v' (graph.suc source) *)
%     end) =
%   sig
%     (*@ predicate has_path (v1 v2 : G.V.t) (g : G.gt) *)

%     val check_path : G.gt -> G.V.t -> G.V.t -> bool
%     (*@ r = check_path graph v1 v2
%           requires mem v1 graph.G.dom
%           ensures r <-> has_path v1 v2 graph *)
%   end
% \end{gospelsmall}
%
\begin{ocamlenv}
  module Check
    (G : sig
      module V : COMPARABLE
      type gt
      val successors : gt -> V.t -> V.t list
    end) =
  sig
    val check_path : G.gt -> G.V.t -> G.V.t -> bool
  end
\end{ocamlenv}
The implementation of this algorithm follows a classical approach.  To
check if there is a path between two nodes, \of{v1} and \of{v2}, we
employ a Breadth First Search starting at \gosp{v1} and, during the
search, test if the visiting node is \gosp{v2}. This is implemented in
\ocaml as follows:
\begin{ocamlsmall}
  module HV = Hashtbl.Make(G.V) (* hash table for vertices *)

  let check_path graph v1 v2 =
    let marked = HV.create 97 in
    let [@ghost] visited = HV.create 97 in
    let q = Queue.create () in
    let rec loop () =
      if Queue.is_empty q then (* exhausted graph and no path found *)
        false
      else
        let v = Queue.pop q in
        if G.V.compare v v2 = 0 then true (* path found *)
        else begin
          HV.add visited v ();
          let rec iter_succ sucs = ... in
          iter_succ (G.successors graph v);
          loop () end in
    HV.add marked v1 ();
    Queue.add v1 q;
    loop ()
\end{ocamlsmall}
The algorithm uses the \of{marked} hash table to tag the vertices
found during traversal, but not yet fully explored (\emph{i.e.}, still
waiting in the traversal queue). This is the role of the auxiliary
\emph{ghost} table \of{visited}, which stores all the vertices that
have been popped from queue~\of{q}. It is worth noting that table
\of{visited} is only used to ease the specification and proof process,
having no runtime
implications~\cite{DBLP:journals/fmsd/FilliatreGP16}.

The core of the algorithm lies in the \gosp{loop} function. An empty
queue implies that we have traversed the graph and were unable to find
\gosp{v2}. However, if the popped element of the queue,~\of{v}, is
equal to \gosp{v2} % , with regards to the vertex compare function,
then we found a path.
%
% \begin{ocamlsmallnc}
%   let rec loop () =
%     if Queue.is_empty q then (* exhausted graph and no path found *)
%       false
%     else
%       let v = Queue.pop q in
%       if G.V.compare v v2 = 0 then true (* found path *)
%       else ...
% \end{ocamlsmallnc}
%
  %% One subtle difference between the presented implementation
  %% and the one present in the \ocamlgraph library is how we iterate over
  %% the list of successors of some node. The library defines a set of
  %% higher-order iteration patterns, such as \gosp{iter_succ}.
  %% Currently, there is no elegant way of specifying higher-order
  %% iteration in \gospellang, which forces us to inline the call to
  %% this function. %This was moved from the previous paragraph, makes more
  %% % sense here i think. Ion: I Agree, thanks!
%
If~\of{v} is not the destination vertex, then we add its successors
to the queue and mark them in the recursive function \gosp{iter_succ}.

% \begin{ocamlsmallnc}
%    let rec iter_succ = function
%      | [] -> ()
%      | v' :: r ->
%        if not (HV.mem marked v') then begin
%          HV.add marked v' () ;
%          Queue.add v' q end;
%          iter_succ (prefix @ [v']) r in
%     HV.add visited v ();
%     let sucs = G.successors graph v in
%     iter_succ [] sucs;
%     loop ()
% \end{ocamlsmallnc}

The complete \ocaml implementation is provided in the companion artifact.

\subsection{Dynamic Analysis of Auxiliary Data Structures, in \ortac}
\label{sec:dynam-analys-auxil}

\begin{figure}[t]
\begin{gospelsmall}
  type (!'a, !'b) t
  (*@ mutable model contents : ('a * 'b) list *)

  val create : ?random: bool -> int -> ('a, 'b) t
  (*@ h = create ?random size
      ensures h.contents = [] *)

  val add : ('a, 'b) t -> 'a -> 'b -> unit
  (*@ add h k v
      modifies h
      ensures h.contents = (k, v) :: old h.contents *)

  val mem : ('a, 'b) t -> 'a -> bool
  (*@ b = mem h k
      ensures b = List.mem k (List.map fst h.contents) *)
\end{gospelsmall}
\caption{Excerpt of the \gosp{Hashtbl} module, specified using \gospellang.}
\label{fig:hashtbl-gospel}
\end{figure}

\mario{Uma frase inicial para dizer que seria o mais natural e tentador tentar
  aplicar \ortac a toda a implementação do functor \gosp{Check}.}

To smoothly tackle the verification of the \of{check_path}
implementation, we chose to dynamically analyze the used auxiliary
data structures. In particular, we analyze the queue and hash table
structures, provided by the \gosp{Queue} and \of{Hashtbl} module from
the \ocaml standard library.

To dynamically analyze the \of{Queue} module, we revisit the
specification depicted in \Cref{fig:queue}. However, for the sake of
the specification being completely executable, hence usable by \ortac,
we must perform some changes to such a specification. Specifically, we
must split the logic of the postcondition of \gosp{pop} into two
clauses: one of the form \of{t.elems = ...}, which captures the state
of the model when the function returns; another to assert how the
returned value relates to the model. Finally, the \stm plugin is only
able to analyze functions featuring a single argument of type %
\of{'a t}, the type of queues, hence we remove the \of{transfer}
function. The adapted \texttt{Queue} specification (equivalent to the
one in \Cref{fig:queue}, used by \ortac, is given in
Appendix~\ref{sec:queue-spec-ortac}.

To analyze the \of{Hashtbl} module in \ortac, we use the \gospellang
specification and hash table operations depicted in
~\Cref{fig:hashtbl-gospel}. The hash table type is defined as %
\gosp{(!'a, !'b) t}, meaning it represents a hash table from keys of
type~\gosp{'a} bound to values of type~\gosp{'b} (the~\gosp{!} symbol
is used in \ocaml type-checker to track variance and covariance of
type parameters). From a logical point of view, an hash table is
modeled as an association list (field \of{contents}), \emph{i.e.}, the
pair \of{(k, v)} is in the model if and only if key~\of{k} is bound to
value~\gosp{v} in the table. Given such model type, the supplied
\gospellang specification for three hash table operations (the ones
used in the \of{check_path} implementation) is the expected one. The
complete specification of the \of{Hashtbl} module is actually part of
the \stm plugin test-suite and is publicly
available\footnote{\url{https://github.com/ocaml-gospel/ortac/blob/main/plugins/qcheck-stm/test/hashtbl.mli}}.

% We briefly explain the specification of the three hash table functions
% declared in ~\Cref{fig:hashtbl-gospel}. These are the three functions
% from the \of{Hashtbl} module we use in the implementation of the
% \of{check_path} function\footnote{Complete specification of the
%   Hashtbl module
%   \url{https://github.com/ocaml-gospel/ortac/blob/main/plugins/qcheck-stm/test/hashtbl.mli}}.
% The \of{create} operation simply returns an empty hash table with
% initial capacity equal to \of{size}. The optional argument
% \of{random}, which is used to control whether the table always uses a
% fixed hash function, is ignored in the specification. Function
% \of{add} inserts a biding for value~\of{v} to key~\of{k} to
% table~\of{h}, via a side-effect to modify the table in-place. Finally,
% \of{mem} is a pure function with no side-effects which simply asserts
% whether key~\of{k} is bound in table~\of{h}. From a specification
% point of view, we first strip way the values from the \of{contents}
% model, converting it into a list of keys using \of{List.map}, and then
% check if~\of{k} is in that list using \of{List.mem}.

We feed the \gosp{Queue} and \of{Hashtbl} specifications to \ortac,
which uses the \stm plugin to generate an instrumented \ocaml wrapper
for random calls on those data structures operations. As candidate
implementations, for both modules, we use the actual implementations
provided by the \ocaml standard library. \ortac successfully checks
such implementations, meaning the randomly generated traces of
function calls do not violate the expected behavior expressed in the
\gospellang specifications.

\subsection{Path Checking Proof, in \cameleer}
\label{sec:path-checking-proof}

To conduct a proof of the \of{check_path} function, we must first
provide a specification to the \gosp{Check} functor, which means we
also need to annotate the module it receives as an argument. This is
as follows:
\begin{gospelsmall}
  type gt
  (*@ model dom: V.t fset
      model succ: V.t -> V.t fset
      with x invariant forall v1, v2.
      mem v1 x.dom -> mem v2 (x.succ v1) -> mem v2 x.dom *)

  val successors : gt -> V.t -> V.t list
  (*@ l = successors graph source
      requires mem source graph.dom
      ensures forall v'. List.mem v' l <-> Fset.mem v' (g.succ source) *)
\end{gospelsmall}
First, we attach two \of{model}s to the \gosp{gt} type: the set of
vertices in the graph, \of{dom}, as well as a successor function
\gosp{succ} that returns all the finite set of successors of a given
vertex. Additionally, we supply a type invariant that states the
\of{dom} set is closed under the \of{succ} function. Finally, the
specification of \of{successor} function states it returns the
complete list of successors of node \of{source}.

We can specify the actual \of{check_path} function. In this, this
function implements an algorithm that decides whether there is a
sequence of edges that allows one to travel from \gosp{v1} to
\gosp{v2}. Such a property is captured by the following \gospellang
predicates:
\begin{gospelonlysmall}
  (*@ predicate is_path (v1 v2: G.V.t) (l: G.V.t seq) (g: G.gt) =
        let len = Seq.length l in
        if len = 0 then v1 = v2 else
          edge v1 l[0] g && l[len - 1] = v2 && mem v1 g.G.dom &&
          forall i. 0 <= i < len - 1 -> edge l[i] l[i+1] g *)

  (*@ predicate has_path (v1 v2 : G.V.t) (g : G.gt) =
        exists_ p. is_path v1 v2 p g *)
\end{gospelonlysmall}
Here, \of{seq} stands for the type of finite mathematical sequences
provided in the \gospellang standard library. The specification of
\of{check_path} is, now, rather intuitive:
\begin{gospelsmall}
  val check_path : G.gt -> G.V.t -> G.V.t -> bool
  (*@ b = check_path g v1 v2
      requires mem v1 graph.G.dom
      ensures b <-> has_path g v1 v2 *)
\end{gospelsmall}
Note that we require~\of{v1} to be a vertex in the graph, since our
implementation maintains the invariant that marked nodes are a
subset of the graph domain.
%
%\begin{gospelonlysmall}
%  invariant subset marked.HV.dom graph.G.dom
%\end{gospelonlysmall}
%
\begin{figure}[t]
  \centering
  \includegraphics[width=0.6\linewidth]{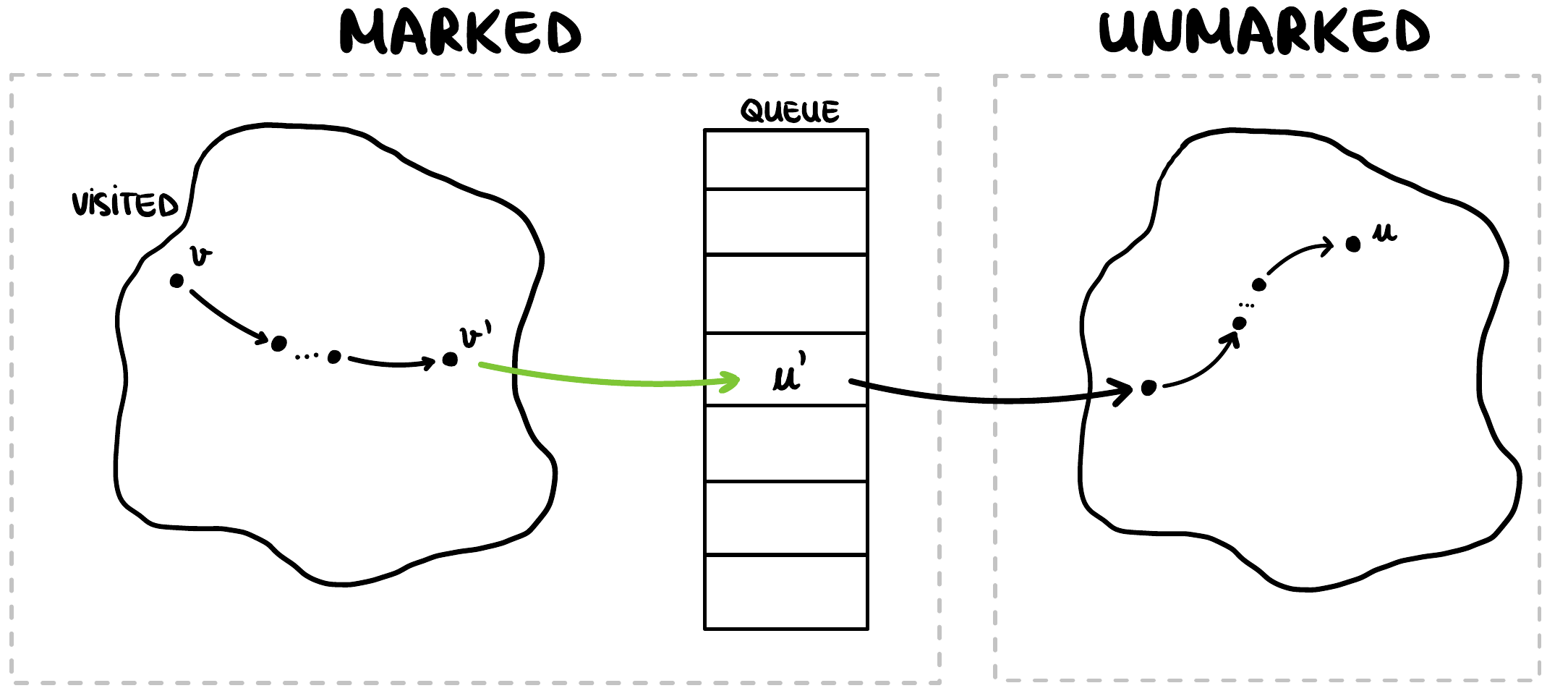}
  \caption{Visual representation of the \gosp{intermediate_value}
    lemma.}
  \label{fig:iter-val}
\end{figure}

Having defined a \gospellang specification for the \of{Check} functor
and the \of{check_path} function, one might be tempted to apply \ortac
to dynamically test the whole \ocaml implementation. However, the
definition of the \of{has_path} predicate prevents us from doing so:
general use of existential quantification does not fall into the
\gospellang's executable fragment.

% Though we universally quantified the predicate \gosp{has_path} in the interface,
% we should refine it, giving it a body, as an existential quantifier of a path
% \gosp{p} between \gosp{v1} and \gosp{v2} in graph \gosp{g}. The path is modeled
% as a sequence with an edge between \gosp{v1} and the first element of the
% sequence; the last element is \gosp{v2}, all while the sequence is a continuous
% linking of edges of the graph.

% \begin{gospelonlysmall}
%   (*@ predicate is_path (v1 v2: G.V.t) (l : G.V.t seq)
%                         (g : G.gt) =
%         let len = Seq.length l in
%         if len = 0 then v1 = v2 else
%           edge v1 l[0] g && l[len - 1] = v2 && Fset.mem v1 g.G.dom &&
%           forall i : int. 0 <= i < len - 1 -> edge l[i] l[i+1] g *)

%   (*@ predicate has_path (v1 v2 : G.V.t) (g : G.gt) =
%         exists_ p. is_path v1 v2 p g *)
% \end{gospelonlysmall}

  %% \subsection{Intermediate Value}

The complete \gospellang specification of the inner functions in the
\of{check_path} definition is given in the companion artifact. We
highlight here only the main invariants of the \of{loop} function that
allows one to prove the \emph{correctness} and \emph{completeness} of
the algorithm.

Correctness means that if \of{check_path} returns \of{true}, there is
indeed a path between \of{v1} and \of{v2} in the graph. This is
achieved by maintaining the invariant that there is always a path
from~\of{v1} to a marked vertex, which includes all vertices in the
queue and the \of{visited} table. This is expressed in \gospellang as
follows:
\begin{gospelonlysmall}
  invariant forall v. mem v marked.HV.dom -> has_path v1 v graph
\end{gospelonlysmall}
Completeness means that if there is no path, then we have indeed fully
explored the graph. In other words, while the traversal is not
finished, if there is a path from \of{v1} to \of{v2}, then such a path
must go trough an intermediate vertex~\of{w} from the queue. This is
expressed in \gospellang as follows:
\begin{gospelonlysmall2}
  invariant has_path v1 v2 graph ->
            exists_ w. Seq.mem w q.Queue.elems /\ has_path w v2 graph
\end{gospelonlysmall2}
After setting on the correct \gospellang specification, the
\of{check_path} proof is done mostly automatically, using a
combination of SMT solvers. The only part requiring extra human
interaction is in proving the completeness invariant. To do so, one
must provide a classic \emph{intermediate value}
lemma. \Cref{fig:iter-val} depicts a visual representation of such a
statement: if there is a path from a marked node~\of{v} to an unmarked
node~\of{u}, then it must be the case that there is an intermediate
edge crossing the set of marked elements to the set of unmarked
elements. In our case, we need to retrieve the actual bridge between
the \of{visited} set and the queue, colored in green in
\Cref{fig:iter-val}. There is a constructive proof of such a result,
which amounts to a recursive traversal from \of{v} to \of{u}, until
one finds the green arrow. This is encoded in \cameleer as a ghost
function.

\subsection{Proof of Auxiliary Data Structures, in \cfmllang}
\label{sec:proof-auxiliary-data}

\mario{Dizer aqui que vamos cobrir algumas limitações do \cameleer e
  do próprio \ortac. Por exemplo, o \ortac não é capaz de analisar a
  função \of{transfer}.}

\mario{O discurso sobre a \gosp{Queue} está muito bem, mas vamos então
  generalizar para as ambas as estruturas auxiliares: \gosp{Queue} e
  \gosp{Hashtbl}.}

\mario{Referir também a diferença entre specs da parte do \ortac e agora que
  estamos na prova. Na prova já não estamos obrigados a usar apenas modelos
  executáveis.}

\mario{A prova do módulo \gosp{Hashtbl} é importante dizermos que foi feita
  num trabalho ``exterior'' ao nosso, em 2017 pelo François Pottier.}

\mario{Quando dizemos que o Cameleer assume uma spec para a hash-table e
  queue, dar o link para o Github do Cameleer onde está o ficheiro com a
codificação da biblioteca standard OCaml e que, portanto, contém a especificação
destes dois módulos.}

As a final piece of our case study, we can further increase our degree
of confidence in the \ocaml \of{check_path} implementation by actually
proving the implementation of the auxiliary queue and hash table data
structures. Reasoning about such structures is out of scope for
\cameleer, since these are pointer-based structures. We must turn
ourselves into \cfmllang.

% Although the path checking algorithm was successfully proven, we did
% so assuming that the implementation of the \ocaml \gosp{Queue} and
% \gosp{Hashtbl} modules are sound with regards to their respective
% \gospellang specification. In the case of \gosp{Queue}, we assume the
% specification in the \Cref{fig:queue}, and in the case of
% \gosp{Hashtbl}, we assume the specification in the \cameleer enconding
% of the \ocaml standard library. To wrap up our proof, we present our
% results in proving the \gospellang specification for the \gosp{Queue}
% module presented in \Cref{sec:gospel}.

We do not conduct a proof for the \gosp{Hashtbl} module, has this was
already done by François Pottier in
2017~\cite{DBLP:conf/cpp/Pottier17}. As for \of{Queue}, we prove here
that the \ocaml standard library implementation adheres to the
\gospellang specification presented in \Cref{fig:queue}. The type of
queues is defined as follows:
\begin{ocamlenv}
  type 'a cell_contents = { content: 'a; mutable next: 'a cell; }
  and  'a cell = Nil | Cons of 'a cell_contents

  type 'a queue = {
    mutable length: int;
    mutable first: 'a cell;
    mutable last: 'a cell
  }
\end{ocamlenv}
Basically, a queue is built on top of a mutable linked-list whose
entry pointer is stored in field \of{first}, while the last element is
pointed by \of{last}. Maintaining these two pointers is what allows
one to achieve constant-time push, pop, and transfer operations.

To start a \cfmllang proof, we must first supply representation
predicates for our structures. For the case of \of{cell}, this is as
follows:
\begin{cfml}
  Definition Cell A (v: A) (n c: cell_ A) : hprop :=
    \exists cf, [c = Cons cf] \* (cf ~~~> `{ content' := v; next' := n }).

  Fixpoint Cell_Seg {A} (L: list A) (to from: cell_ A) : hprop :=
    match L with
    | nil => [to = from]
    | x :: L' => \exists n, (from ~> Cell x n) \* (n ~> Cell_Seg L' to)
    end.
\end{cfml}
The \gosp{Cell} predicate claims ownership of a single heap-allocated
cell~\of{c}, whose \of{content} is~\of{v} and the \of{next} element is
cell~\of{n}. This definition uses notation \gosp{cf ~~~> ... }, which stands for
a cell~\of{cf} that owns the fields of a record. The \of{Cell_Seg} predicate
captures the common notion of a \emph{list segment}. It is worth noting that
with such a representation, if %
\of{c1 ~> Cell_Seg L c2} holds for some cells~\of{c1} and~\of{c2} and a logical
list~\of{L}, then the predicate does not own pointer~\of{c2}.  We equip the
\gosp{queue} type with a representation predicate, as follows:
\begin{cfml}
  Definition Queue A (L: list A) (q: loc) : hprop :=
    \exists (cf cl: cell_ A),
    (q ~~~> `{ length' := length L; first' := cf; last' := cl }) \*
      If L = nil
        then [cf = Nil] \* [cl = Nil]
        else \exists x L', [L = L' & x] \*
             (cf ~> Cell_Seg L' cl) \* (cl ~> Cell x Nil).
\end{cfml}
This predicate claims ownership of all the elements the \gosp{first}
list contains, except the last one, and claims ownership of the
\of{last} cell, whose contents is equal to the last element of the
logical list~\of{L}.

The above representation predicates are the building blocks to provide
specification to queue operations and conduct interactive proof using
\cfmllang. The complete proof of the \of{Queue} module, \emph{i.e.},
representation predicates, auxiliary lemmas, and specification
verification for each function, is provided in the companion artifact.

% why3 replay path  108.66s user 3.67s system 261% cpu 43.007 total

\begin{table}[t]
  \centering
  \begin{tabular}
    {
   >{\raggedright}m{.22\textwidth}|
   >{\centering}m{.12\textwidth}|
   >{\centering}m{.14\textwidth}|c|c}
    Case study & \# VCs & LoC / LoS & Proof Replay (s) & Fully Auto. \\\hline
    \rowcolor{thegray}
    Path checking & & & & \\
    \hspace{1em} \ortac & - & 11  / 42 & - & \check \\
    \hspace{1em} \cameleer & 213 & 50 / 118 & 26.63  & \cross \\
    \rowcolor{thegray}
    Queue & & & &  \\
    \hspace{1em} \ortac & - & 6 / 21 & -  & \check \\
    \hspace{1em} \cfmllang & 4 & 19  / 72 & -  & \cross \\
    \hspace{1em} \cameleer & 47 & 31  / 25 & 1.93  & \check \\
    \rowcolor{thegray}
    Hashtbl & & & & \\
    \hspace{1em} \ortac & - & 32 / 45 & - & \check \\
    \hspace{1em} \cfmllang & 89 & 186 / 1397 & - & \cross \\
  \end{tabular}
  \caption{Summary of the case studies and their respective statistics.}
  \label{figure:case-studies}
\end{table}

\Cref{figure:case-studies} showcases a summary of the case
studies. For the different tools we present the number of
Verifications Conditions, in the case of \cameleer, and number of
Theorems and Lemmas in the case of \cfmllang, the number of lines of
code (LoC) and lines of specification/script (LoS), the proof replay
time in \cameleer (not applicable to \cfmllang and \ortac) and if the
conducted proofs were fully automatic. These measurements were
conducted on a Lenovo Thinkpad X1 Carbon 8th Generation machine,
running Linux kernel 5.4.0-60-generic, 4 Intel 1.80 GHz CPUs, and 16
Gb of RAM. The time displayed is the average of 5 runs, measured using
the \texttt{hyperfine} benchmarking tool.
% The path checking case study, covered by \ortac and \cameleer

\section{Related Work}
\label{sec:related-work}

Given deductive verification and runtime assertion checking are two active
fields of research, the literature on these topics is quite vast. We do not aim
to survey it here. Instead, in this section we focus on verification frameworks
that combine static and dynamic analysis under the same umbrella.

The \textsf{Frama-C}~\cite{DBLP:journals/fac/KirchnerKPSY15} framework
provides a tool set for the certification of C programs, based on
\textsf{ACSL}, the ANSI/ISO-C Specification Language. On top of
\textsf{ACSL}, \textsf{Frama-C} features a modular architecture based
on plugins~\cite{DBLP:journals/corr/Signoles15}. Each of such plugins
implements a different analysis strategy. The \textsf{E-ACSL}
plugin~\cite{DBLP:conf/rv/SignolesKV17} establishes an executable
subset for \textsf{ACSL}, hence it can be used to perform runtime
assertion checking on C programs. Similar to \textsf{Frama-C}, one can
cite \textsf{OpenJML}~\cite{DBLP:conf/ecoop/Cok21}, based on
\textsf{JML} (the Java Modeling Language), and \textsf{SPARK} for the
Ada programming language. Unlike \textsf{ACSL}, however, \textsf{JML}
and \textsf{SPARK} impose specifications to always be executable. The
three mentioned analysis frameworks target only imperative languages,
whereas the \gospellang ecosystem is concerned with the verification
of programs written in \ocaml, a multi-paradigm language.

% Frama-C --> E-ACSL
% OpenJML --> JML --> spec executável

Kosmatov \textit{et al.}~\cite{DBLP:conf/isola/KosmatovMMS16} survey static and
dynamic verification within \whythree, \textsf{Frama-C} and \textsf{SPARK}
2014. Both \textsf{Frama-C} and \textsf{SPARK} conduct deductive verification by
translating an input program, written either in C or Ada, into an equivalent
\whyml program. In the case of deductive verification using \gospellang
specification, we use \cameleer to conduct deductive verification via \whythree,
but we can also interface with \coq, via \cfmllang, when one needs to use
Separation Logic to conduct some proof. Finally, the authors also describe the
use of counterexamples, either generated by SMT solvers or via testing, to debug
proof failures. Adding support for the generation of counterexamples, expressed
as \gospellang terms, would be an interesting addition to the ecosystem.

Achieving full combination of static and dynamic analyses is known to
be an important challenge in the field of formal
methods~\cite{DBLP:conf/fmco/LeavensCCRC02}. Different specification
styles, targeting different back-end tools (\emph{e.g.}, automated
solvers, interactive proof assistants, or execution monitors), and the
question of how to make the two analyses agree on a common semantics,
makes it a non-trivial task to readily combine the two approaches. To
this regards, Maurica \textit{et
  al.}~\cite{DBLP:conf/isola/MauricaCS18} survey the architecture and
design choices of \textsf{Frama-C} and \textsf{OpenJML} towards
approximating runtime assertion checking capabilities to those of
static verification. Within the \gospellang ecosystem, one of our
major goals is to bridge the gap between \ortac and the deductive
verification tools. For instance, we aim at improving the
expressiveness of supported executable \gospellang subset
(\emph{e.g.}, quantification) and incorporating memory analysis
techniques, based on Separation
Logic~\cite{10.1007/978-3-540-78163-9_19}.

\section{Conclusions and Future Work}
\label{sec:concl-future-work}

In this paper, we reported on our experience using dynamic and static analysis
tools to, collaboratively and incrementally, tackle the verification of \ocaml
code. In our certification pipeline, we first apply \ortac, via the
\textsf{QCheck-STM} plugin, to perform property-based state-machine testing of
parts of the program. Then, \cameleer and \cfmllang are used to deductively
verify \ocaml implementations, hence complementing the correctness guarantees
provided by the RAC step. However, while conducting the proof, one can rely on
the dynamically analyzed specification to use, for instance, as the logical
behaviour of auxiliary data structures. We demonstrate how this pipeline works in
practice, by applying it to an \ocaml implementation of a path checking
algorithm in a graph.

We use \gospellang, the \ocaml specification language, as the key
ingredient to achieve collaboration between the two analysis
approaches. \gospellang is not tied to any particular tool or even
verification methodology, hence it is used as a common ground for RAC
and deductive verification of parts of the same piece of \ocaml
code. We believe our proposal for the combined use of analysis tools,
based on user-supplied \gospellang specifications, is an important
contribution towards a wider adoption of formal methods by the \ocaml
community. To integrate well with common development cycles and be
applicable to the verification of industrial-scale systems, a
certification workflow should be flexible enough and offer different
levels of assurance. It is our goal to keep improving our methodology
and tools to analyze more realistic, industrial-size \ocaml case
studies.
We conclude with other avenues of future work that we believe are worth to
explore.

\paragraph{Differences between specification for RAC and Deductive
  Verification.} As presented throughout the paper, \ortac imposes a particular
style for writing \gospellang specification that makes such a specification less
natural when compared to what one writes for a proof. Even if we argue,
informally, that the two sorts of specification are logically equivalent, it
would be interesting to derive a formal argument of such relationship. One
possibility would be to generate a proof script stating specification inclusion
between \gospellang statements used in \ortac and those used in \cameleer or
\cfmllang.

\paragraph{RAC when deductive verification fails.} In our methodology, when
doing a proof, we rely upon a \gospellang specification that has been tested in
\ortac. We showcase this approach in our example of the path checking algorithm,
where the \cameleer proof is conducted against the \texttt{Queue} and
\texttt{Hashtbl} specifications checked by \ortac. As such, we explore the
results of dynamic analysis during the static phase. Allowing for the opposite
direction, \emph{i.e.}, going from deductive verification into RAC, would also
be of interest. One possibility is to apply RAC tools when a proof does not
succeed. For instance, when one fails to prove a loop invariant it could
generate an executable version of such an invariant (a \emph{monitor}) to
dynamically analyze it. The SPARK 2014 toolset~\cite{McCormick_Chapin_2015} is a
successful example of this approach. In the case of the \gospellang ecosystem,
we would likely need to extend \ortac to deal not only with \ocaml interfaces,
but also with implementation files.

% \paragraph{The last mile: verified compilation.}

\subsubsection{Acknowledgments.} We thank Ana Ribeiro for her support
in designing Figures~\ref{fig:pipeline} and~\ref{fig:iter-val}. We
thank the ISoLA 2024 anonymous reviewers. Their comments and
suggestions have greatly improved the presentation of this paper.
This work is partly supported by Agence Nationale de la Recherche
(ANR) grant ANR-22-CE48-0013-01 (GOSPEL) and NOVA LINCS
ref. UIDB/04516/2020 (\url{https://doi.org/10.54499/UIDB/04516/2020})
and ref. UIDP/04516/2020
(\url{https://doi.org/10.54499/UIDP/04516/2020}) with the financial
support of FCT.IP.

\mario{Verified compilation: \ocaml into CakeML?}

\mario{Agora estou na dúvida. Será que falar de compilação certificada faz
  sentido no nosso contexto?}

\bibliographystyle{splncs04}
\bibliography{bibliography}

\newpage

\appendix

\section{COMPARABLE Module Signature}
\label{sec:vertex}

\begin{ocamlsmall}
  module type COMPARABLE = sig
    type t
    val compare : t -> t -> int
    val hash : t -> int
    val equal : t -> t -> bool
  end
\end{ocamlsmall}

\section{Queue Specification for \ortac}
\label{sec:queue-spec-ortac}

\begin{gospelsmall}
type 'a t
(*@ mutable model elems : 'a list *)

val create : unit -> 'a t
(*@ t = create ()
    ensures t.elems = [] *)

val is_empty : 'a t -> bool
(*@ b = is_empty t
    ensures b = t.elems = [] *)

val push : 'a -> 'a t -> unit
(*@ push x t
    modifies t.elems
    ensures t.elems = (old t.elems) @ [x] *)

exception Empty

val pop : 'a t -> 'a
(*@ x = pop t
    modifies t.elems
    ensures t.elems = if old t.elems = []
                         then []
                         else List.tl (old t.elems)
    ensures if old t.elems = [] then false
            else x = List.hd (old t.elems)
    raises Empty -> old t.elems = [] = t.elems *)
\end{gospelsmall}
\end{document}

%% file: config.tex
\usepackage[T1]{fontenc}
\usepackage{graphicx}
\usepackage{subcaption}
\usepackage{xspace}
\usepackage{amsmath}
\usepackage{listings}
\usepackage{xcolor}
\usepackage{bold-extra}
\usepackage{listings}
\usepackage{tikz}
\usepackage{todonotes}
\usepackage[skins]{tcolorbox}
\usepackage{latexsym}
\usepackage[os=win, mackeys=symbols]{menukeys}

\definecolor{codemarkcolor}{RGB}{178, 190, 181} % Define the color (you can adjust the RGB values)

\renewcommand{\check}{
 \raisebox{-.25\height}{\includegraphics[width=1em,height=1em]{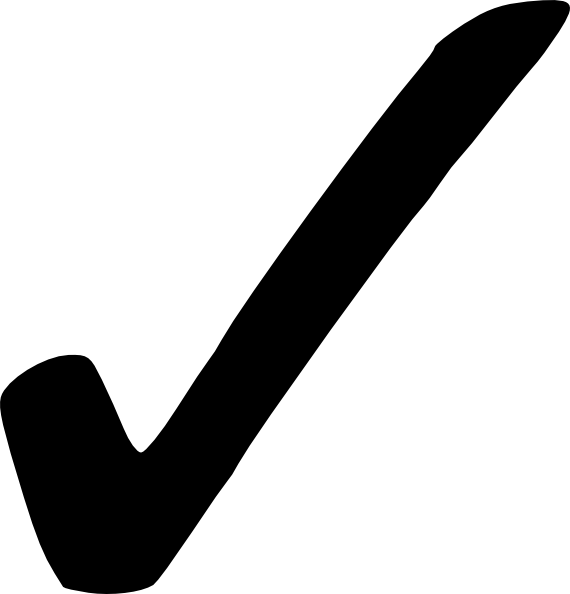}}}

\newcommand{\cross}{
 \raisebox{-.25\height}{\includegraphics[width=1em,height=1em]{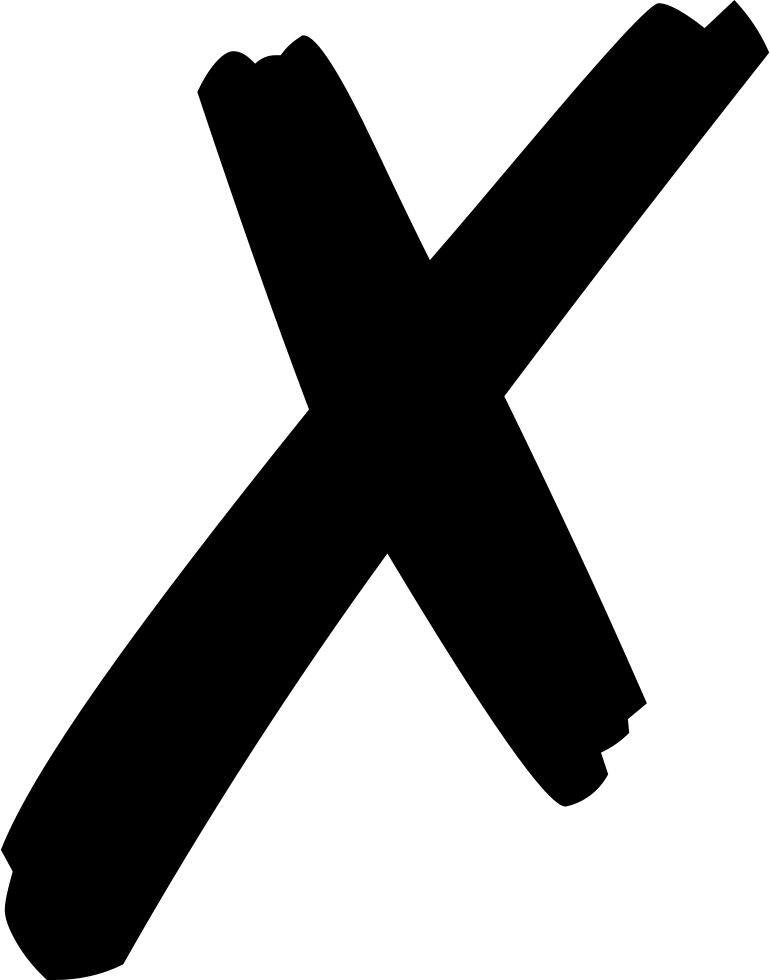}}}

\usepackage{gospel}
\DeclareCaptionFormat{listing}{\hfill#3}

%% file: mymacros.tex
\usepackage{color}

\newcommand{\whyml}{\textsf{WhyML}\xspace}

\newcommand{\ocaml}{\textsf{OCaml}\xspace}
\newcommand{\ocamlgraph}{\textsf{OCamlGraph}\xspace}

\newcommand{\coq}{\textsf{Coq}\xspace}

\newcommand{\cfmllang}{\textsf{CFML}\xspace}

\newcommand{\gospellang}{\textsf{Gospel}\xspace}
\newcommand{\cameleer}{\textsf{Cameleer}\xspace}
\newcommand{\whythree}{\textsf{Why3}\xspace}

\newcommand{\ortac}{\textsf{Ortac}\xspace}
\newcommand{\stm}{\textsf{QCheck-STM}\xspace}
\definecolor{thegray}{rgb}{0.9,0.9,0.9}
\definecolor{colorspec}{rgb}{0,0,0.797}
\definecolor{thered}{rgb}{0.797,0,0}
\definecolor{darkgreen}{rgb}{0.797,0,0}
\definecolor{theblue}{rgb}{0,0,0.797}
\definecolor{darkgray}{rgb}{0.8477,0.8477,0.8477}
\definecolor{ocaml-bg}{rgb}{0.9,0.9,0.9}
\definecolor{thegraygray}{rgb}{0.5,0.5,0.5}